\def\imat {\rm i} 

\documentstyle[twocolumn,floats,epsf,ifthen,aps,prb]{revtex} 

\begin{document} 

%\preprint{draft} 
 
%\draft 

\title{Quantum feedback control of a solid-state qubit} 

\author{Rusko Ruskov\footnote{On leave of absence from 
Institute of Nuclear Research and Nuclear Energy, Sofia BG-1784, Bulgaria} 
and Alexander N. Korotkov\footnote{Electronic mail: korotkov@ee.ucr.edu}} 
\address{ 
Department of Electrical Engineering, University of California, 
Riverside, CA 92521-0204. 
} 
\date{\today} 
 
\maketitle 
 
\begin{abstract} 
We have studied theoretically the basic operation of a quantum feedback loop 
designed to maintain a desired phase of quantum coherent oscillations 
in a single solid-state qubit. The degree of oscillations synchronization with 
external harmonic signal is calculated as a function of feedback strength, 
taking into account available bandwidth and coupling to environment. 
 The feedback can efficiently suppress the dephasing of oscillations 
if the qubit coupling to the detector 
is stronger than coupling to environment.
\end{abstract} 
%\pacs{PACS numbers: }
%\pacs{73.23.-b; 03.65.Bz}

%\newpage
\narrowtext 
 
%\vspace{1ex} 
\vspace{0.6cm} 

        The principle of feedback control is used in a wide variety
of physical and engineering problems. In particular, it can be applied
in a straightforward way to tune the oscillation phase of a harmonic 
oscillator in order to achieve a desired synchronization with some 
reference oscillator. An intriguing and fundamental question is whether
continuous feedback can be used to control quantum systems; for instance, 
if it is possible or not to tune the phase of quantum coherent (Rabi)
oscillations in a qubit (two-level system). 

        At first sight the quantum feedback seems to be impossible
because according to the ``orthodox'' collapse postulate \cite{Neumann}  
the quantum state is abruptly destroyed by the act of measurement. 
However, as was shown two decades ago, in particular by Leggett,  
\cite{Caldeira} in a typical solid-state setup the collapse of a 
qubit state should be considered as a continuous process rather
than as instantaneous event. The reason is typically weak coupling
between the quantum system and the detector and also the finite noise
of the detector, so that it takes some time until acceptable 
signal-to-noise ratio is reached and the measurement can be regarded  
as completed. 

        While the Leggett's theory as well as the majority of similar 
approaches can describe only {\it ensembles} of quantum systems, 
the theory describing the gradual collapse of a {\it single} solid-state 
qubit was developed only recently.\cite{Kor-PRB,Kor-meas2,Goan} 
(A similar problem in quantum optics was solved much earlier -- see, e.g. 
Refs.\ \cite{Carmichael,Wiseman} and references in \cite{Kor-meas2}.) 
Basically, the theory says that the evolution of a single quantum system 
due to continuous measurement is governed by the information continuously 
acquired from the detector. Similarly to classical probability,
the Bayes formula \cite{Feller} which naturally takes into account incomplete 
information from the detector, can still be applied to the density matrix
of the measured quantum system; thus the formalism is called 
Bayesian.\cite{Kor-PRB} 

        In case of a poor detector the extra noise acting back onto the input 
disturbs the measured system stronger than the limit determined by 
the uncertainty principle; this leads to gradual decoherence of the 
measured system. In contrast, when measured with a good (quantum-limited) 
detector, the quantum system does not loose the coherence (even though
the quantum state evolves randomly); moreover, 
its density matrix can be gradually purified\cite{Kor-PRB} 
that basically means acquiring as much information about the system 
as permitted by quantum mechanics. 

        Since the Bayesian formalism allows us to monitor the continuous
evolution of a quantum system in a process of measurement, this naturally 
gives rise to a possibility of continuous feedback control of a quantum
system. In this paper we will study the operation of a feedback loop 
proposed in Ref. \cite{Kor-meas2} and designed to maintain a desired 
phase of quantum coherent oscillations in a solid-state qubit. 
(Quantum feedback in quantum optics has been proposed and studied earlier 
-- see, e.g., Refs. 
\cite{Wiseman,Caves,Wiseman-94,Tombesi,Hofmann,Wang,Doherty,Doherty-3}.)  
In particular, we will study the dependence of the loop operation
on the feedback strength, available bandwidth, 
and dephasing due to environment. 

\begin{figure} 
\centerline{
\epsfxsize=2.7in 
%\hspace{0.3cm}
\epsfbox{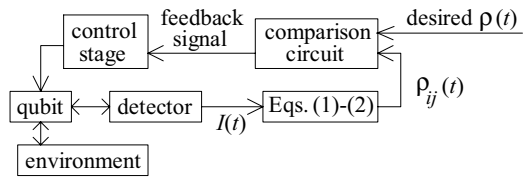}
%\epsfbox{qq.eps} 
} 
\vspace{0.4cm} 
\caption{ Schematic of the quantum feedback loop maintaining the 
quantum oscillations in a qubit. 
 }
\label{schematic}\end{figure}

        As an example of the measurement setup (Fig.\ \ref{schematic}) 
we consider a qubit represented by a single electron in a  
double quantum dot (DQD), the location of which is measured by 
a quantum point contact (QPC) nearby in a way used in Ref.\ \cite{Buks}. 
If the electron is in the dot 2 (state $|2\rangle$) which is closer  
to QPC than dot 1, then the QPC tunnel barrier is higher and so 
the average current $I_2$ through QPC is smaller than the average current 
$I_1$ corresponding to the electron in the dot 1 (state $|1\rangle$). 
Consequently, from the QPC current one gets information about the 
electron location. 
We consider a realistic case of weak response, 
$\Delta I\equiv I_1-I_2 \ll I_0\equiv (I_1+I_2)/2$.  
In this case the measurement time $S_I/2(\Delta I)^2$, which is necessary
to achieve signal-to-noise ratio equal to 1 (here $S_I$ is the shot 
noise of the QPC current), is much larger than $e/I_0$, so the QPC current 
$I(t)$ is continuous on the measurement timescale and we do not need 
to consider individual tunneling events in QPC.

        The evolution of the qubit density matrix $\rho$ during the 
measurement process is described within the Bayesian formalism 
by equations \cite{Kor-PRB,Kor-meas2}  
        \begin{eqnarray} 
\dot{\rho}_{11}= &&  -\dot{\rho}_{22}=  
-2\,\frac{H}{\hbar}\,\mbox{Im}\,\rho_{12}
         +\rho_{11}\rho_{22}\, \frac{2\Delta I}{S_I}\, [I(t)-I_0],  
        \label{Bayes1}\\ 
 {\dot\rho}_{12}= &&  \imat\, \frac{\varepsilon}{\hbar }\,\rho_{12}+ 
        \imat \, \frac{H}{\hbar } \, (\rho_{11}-\rho_{22})
        \nonumber \\ 
&& {}  -( \rho_{11}-  \rho_{22})  \frac{\Delta I}{S_I} \, 
[I(t)-I_0]\, \rho_{12} -\gamma \rho_{12} \, ,      
        \label{Bayes2} 
        \end{eqnarray}
where $\varepsilon$ and $H$ are, respectively, the energy asymmetry 
and tunneling strength of the qubit [the qubit Hamiltonian is 
${\cal H}_{qb}=(\varepsilon /2) (c_2^\dagger  c_2 - c_1^\dagger c_1) 
+ H (c_1^\dagger c_2 +c_2^\dagger c_1)$], and $\gamma =\gamma_d+\gamma_e$ 
is the dephasing rate due to the detector nonideality ($\gamma_d$) and coupling
with the environment ($\gamma_e$).\cite{decay} Theoretically, $\gamma_d =0$ 
when qubit is measured by a QPC; however, if instead of QPC we use a 
single-electron transistor (SET), then dephasing $\gamma_d$ is usually quite 
significant \cite{Kor-meas2,Shnirman} (except the case when the SET operates
in a cotunneling regime \cite{VanDenBrink,Averin-cotun}). 

  Notice that the ensemble dephasing rate $\Gamma =\gamma +(\Delta I)^2/4S_I$
is larger than $\gamma$ because of different evolution of the ensemble
members due to random $I(t)$. Individual realizations can be simulated  
using the formula \cite{Kor-meas2} 
                \begin{equation}
I(t) -I_0 =  (\rho_{11} -\rho_{22}) \, \Delta I/2 +\xi (t) , 
        \label{I(t)}\end{equation}
where $\xi (t)$ is the pure white noise with spectral density $S_\xi =S_I$. 
If Eqs.\ (\ref{Bayes1})--(\ref{Bayes2}) are averaged over $\xi(t)$
(we use Stratonovich definition for stochastic differential equations),
then we get usual ensemble-averaged equations for qubit evolution 
(terms proportional to $\Delta I$ will disappear and $\gamma$ will be
replaced by $\Gamma$). 

        It is natural to characterize the effect of extra dephasing 
$\gamma_d$
by the detector ideality (efficiency) 
$\eta \equiv 1/[1+ \gamma_d 4S_I /(\Delta I)^2]$. 
One can show\cite{Kor-meas2,Averin} that $\eta = (\hbar/2\epsilon_d)^2$ 
where $\epsilon_d$ is the total energy sensitivity of the detector 
[$\epsilon_d \equiv (\epsilon_i\epsilon_o)^{1/2}$ where $\epsilon_o$ is
the usual (output) energy sensitivity and $\epsilon_i$ is a similar
quantity characterizing backaction to the input]. 
 So, an ideal case $\eta =1$ corresponds to a detector 
with quantum-limited sensitivity. 

        To realize a feedback loop (Fig.\ \ref{schematic}), we can monitor 
the qubit evolution using the detector current $I(t)$ plugged into Eqs.\
(\ref{Bayes1})--(\ref{Bayes2}). Then the qubit state is compared 
with the desired state, and the difference signal is used to
control the qubit parameters $H$ and/or $\varepsilon$. In the example
studied in this paper the feedback loop is designed to stabilize
the quantum oscillations of the state of a symmetric qubit ($\varepsilon =0$),
so the desired evolution is $\rho_{11}(t)=1-\rho_{22}(t)=
[1+\cos (\Omega t)]/2$, $\rho_{12}(t)=\rho_{21}^*(t)=
\imat \sin (\Omega t)/2$, where the frequency is 
$\Omega = (4H^2+\varepsilon^2)^{1/2}/\hbar = 2H/\hbar$. 
As a difference (``error'') signal we use the phase difference $\Delta \phi$
($|\Delta \phi |<\pi$)
between the desired value $\phi_0(t)=\Omega t\,\,\, (\mbox{mod}\, 2\pi)$ 
and the monitored value 
$\phi (t) \equiv \arctan \{ 2\, \mbox{Im} \rho_{12}(t)/
[\rho_{11}(t)-\rho_{22}(t)]\}$. This difference is used to control 
the qubit parameter $H$ (changing the barrier height of DQD);   
here we study a linear control: $H_{fb}=(1-F\times \Delta \phi)H$, 
where $F$ is the dimensionless feedback factor.\cite{feedback-Doherty}

In this paper we neglect
additional time delay\cite{Kor-meas2} in the feedback network, however, 
we take into account the finite bandwidth of a line carrying detector
current (that is a critical parameter for a possible experiment). 
More specifically, we average the current $I(t)$ with a rectangular
window of duration $\tau_a$, 
$I_a(t) \equiv \tau_a^{-1} \int_{t-\tau_a}^t I(t')dt'$, 
before plugging it into Eqs.\ (\ref{Bayes1})--(\ref{Bayes2}), so 
that the ``available'' density matrix $\rho_a(t)$ differs from the
``true'' density matrix $\rho (t)$. 
Also, to compensate for the corresponding implicit time delay, 
we use $\Delta \phi = \phi_a - \Omega (t-\kappa\tau_a)$ with
$\kappa =1/2$ (we tried various $\kappa$ and found that $\kappa =1/2$
provides the best operation of the feedback loop).

\begin{figure} 
\centerline{ 
\epsfxsize=2.8in 
%\hspace{0.3cm}
\epsfbox{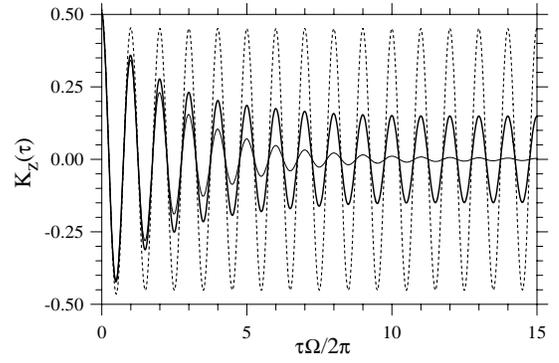}
%\epsfbox{qq.eps} 
} 
\vspace{0.4cm} 
\caption{Correlation function $K_z(\tau)$ of the qubit quantum oscillations 
for ${\cal C}=1$ and feedback factors $F=0$ (thin solid line), 
0.05 (thick solid line), and 0.5 (dashed line). Nondecaying oscillations
are due to synchronization by the feedback. 
 }
\label{K(tau)}\end{figure}

        Let us start with the case of ideal detector, $\eta =1$, 
absence of extra environment, $\gamma_e=0$, and infinite bandwidth, 
$\tau_a=0$. 
Figure \ref{K(tau)} shows numerically calculated correlation function 
$K_z (\tau) \equiv \langle z(t+\tau) z(t)\rangle$ where 
$z\equiv \rho_{11}-\rho_{22}$, for several feedback factors:
$F=0,$ 0.05, and 0.5. 
The curves are obtained using Monte Carlo simulation\cite{Kor-PRB,Kor-meas2} 
of the measurement process for moderately weak coupling between the qubit 
and detector: ${\cal C}\equiv \hbar (\Delta I)^2/S_IH=1$ (notice
that the $Q$-factor of oscillations\cite{Kor-sp} is equal to $8/{\cal C}$,
so ${\cal C}=1$ is still a weak coupling). In absence of feedback 
($F=0$) the correlation function decays to zero (Fig.\ \ref{K(tau)}) 
while for finite feedback factor the correlations remain for indefinitely 
long time (of course, assuming perfect reference oscillator which determines 
the desired evolution). The nondecaying correlations show that the quantum 
feedback loop really provides the synchronization of quantum oscillations.
	The degree of synchronization depends on the feedback factor $F$.
One can see that for a moderate value of $F=0.5$ the synchronization is
already very good [the ideal case would be 
$K_z(\tau )= \cos (\Omega \tau )/2$]. 

\begin{figure} 
\centerline{ 
\epsfxsize=3.0in 
%\hspace{0.3cm}
\epsfbox{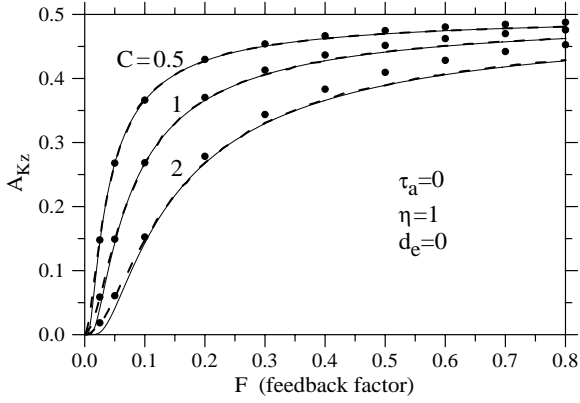}
} 
\vspace{0.4cm} 
\caption{Dots: asymptotic amplitude $A_{Kz}$ of $K_{z}(\tau )$ oscillations
as a function of feedback factor $F$ for several couplings with the detector,
${\cal C}=0.5$, 1, and 2. Solid lines: analytical approximation 
$A_{Kz}=\exp (-{\cal C}/16F)/2$. Dashed lines: corresponding numerical results
for $D^2/2$. 
 }
\label{fig-new}\end{figure}

	For analytical analysis we take into account that 
in the ideal case $\gamma_d=\gamma_e=0$ the qubit state is pure,  
\cite{Kor-meas2} and using Eqs.\ (\ref{Bayes1})--(\ref{I(t)}) start with 
the equation 
	\begin{equation}
\frac{d}{dt}\, \Delta \phi = -\sin \phi \, \frac{\Delta I}{S_I} 
\left( \frac{\Delta I}{2}\, \cos\phi +\xi \right) -
\frac{2FH}{\hbar}\,\Delta\phi ,
	\end{equation}
which assumes the absence of $2\pi$ phase slips (good or moderate 
synchronization). For weak coupling (${\cal C}/8 \ll 1$) 
we can neglect the first term in parentheses and average the random
term over $\sin \phi$ assuming almost harmonic evolution that leads
to the simplified equation
	\begin{equation}
\frac{d}{dt}\, \Delta \phi = \tilde{\xi} 
- \frac{2FH}{\hbar}\,\Delta\phi , 
	\end{equation}
where $\tilde\xi (t)$ is the white noise with spectral density 
$S_{\tilde\xi}=(\Delta I)^2/2S$. This equation describes a particle 
diffusion in the parabolic potential (we again assume $|\Delta\phi | <\pi$).
The corresponding Fokker-Planck equation has an exact solution
which is used to calculate the correlation function
$K_z(\tau )\approx \langle
\cos [\Delta \phi (t)-\Delta \phi (t+\tau ) ] \rangle \cos \Omega \tau  /2$. 
   In this way we obtain the analytical expression 
        \begin{equation}
K_z(\tau )=\frac{\cos \Omega \tau}{2} \,  \exp \left[
\frac{{\cal C}}{16F} \,
\left( e^{-2FH\tau /\hbar} -1\right) \right] , 
        \label{K_z}\end{equation} 
which fits well the Monte-Carlo results when ${\cal C}/8 \ll 1$  
and ${\cal C}/16F \lesssim 1$ (weak coupling and moderate 
or good synchronization).  
	As an example, 
the dots in Fig.\ \ref{fig-new} show the numerically calculated 
(using the least-mean-square fit) asymptotic amplitude $A_{Kz}$ 
of $K_z(\tau )$ oscillations (at $\tau \rightarrow \infty$) as a function 
of the feedback factor $F$ for three values of the coupling $\cal C$, while
solid lines show the corresponding analytical curves 
$A_{Kz}=\exp (-{\cal C}/16F)/2$.

   The correlation function $K_I(\tau )\equiv \langle I(t+\tau )I(t)\rangle$
of the detector current $I(t)$ is somewhat similar to $K_z(\tau )$, however, 
it also has the decaying contribution\cite{Kor-sp} due to correlation 
$K_{z\xi}$ 
and a $\delta$-function contribution due to the detector noise. The analytical 
result for the same regime as above, 
        \begin{eqnarray}
K_I(\tau )= && \frac{S_I}{2}\, \delta (\tau) +\frac{(\Delta I)^2}{4} \,
\frac{\cos (\Omega \tau )}{2} \, \left(1 +e^{-2FH\tau /\hbar} \right) 
\nonumber \\
&& \times \, \exp \left[ ({\cal C}/16F)\left( e^{-2FH\tau /\hbar}-1\right) 
\right]   ,
        \label{K_I}\end{eqnarray} 
also agrees well with the Monte Carlo results. 

        The spectral density $S_{I}(\omega )$ of the detector current 
can be obtained as a Fourier transform of $K_I(\tau )$. While in absence
of feedback the quantum oscillations in the qubit can provide only a moderate 
peak of $S_I(\omega )$ around frequency $\Omega$ (the peak height 
cannot be larger than 
4 times the noise pedestal\cite{Kor-sp}), the feedback synchronization 
leads to the appearance of a $\delta$-function at the frequency of desired 
oscillations. (In principle the desired frequency can differ a little 
from $\Omega$; however, in this case the performance of the feedback
loop worsens.)

      Besides the correlation function and spectral density, we have  
studied one more characteristic, $D$, of the synchronization degree. 
We define $D$ as the average scalar product of the unity-length vector 
on the Bloch sphere corresponding to the desired state and the vector 
corresponding to the actual state of the qubit. The equivalent definition is 
$D\equiv 2 \langle \mbox{Tr} \rho \rho_d \rangle -1$, where $\rho_d$ is the 
density matrix of the desired pure state.
[The so-called fidelity is equal to either $(D+1)/2$ or $\sqrt{(D+1)/2}$, 
depending on the definition.\cite{Doherty-3}] 
 Perfect synchronization corresponds to $D=1$. 
It is simple to show that in the limit of weak coupling and for 
symmetric distribution of $\Delta \phi$ (unshifted desired frequency), 
$A_{Kz}$ coincides with $D^2/2$. Notice, however, that at moderate
coupling, $D^2/2$ 
(see dashed lines in Fig.\ \ref{fig-new}) is significantly closer 
to the analytical result than $A_{Kz}$.

\begin{figure}
\centerline{
\epsfxsize=3.0in 
%\hspace{0.3cm}
\epsfbox{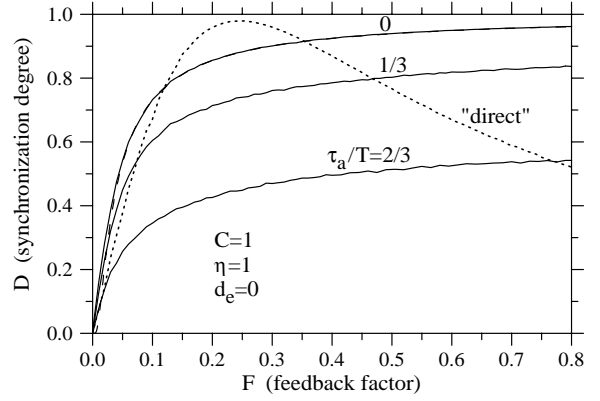}
%\epsfbox{qq.eps} 
} 
\vspace{0.4cm} 
\caption{Synchronization degree $D$ as a function of feedback factor $F$ 
for several values $\tau_a$ of detector signal averaging: $\tau_a/T=0$, 
1/3, and 2/3, where $T=2\pi/\Omega$. Dashed line $D=\exp (-{\cal C}/32F)$
almost coincides with the upper curve. 
 Dotted line corresponds to ``direct'' feedback with $\tau_a= T/10$. 
 }
\label{D(F)}\end{figure}

     Upper solid line in Fig.\ \ref{D(F)} shows the dependence 
of $D$ on the feedback factor $F$ for ${\cal C}=1$ and $\tau_a=0$. 
One can see that $D$ is proportional
to $F$ for small $F$ (``soft'' onset of the synchronization) and $D$ 
is asymptotically approaching 1 at large $F$. 
The analytical result $D = \exp (-{\cal C}/32F)$ 
(dashed line in Fig.\ \ref{D(F)})
is very close to the numerical results at moderate 
and good synchronization. 

      Finite available bandwidth of the detector current $I(t)$ 
(finite averaging time $\tau_a$ in our formalism) worsens the
performance of the quantum feedback loop. The solid lines in Fig.\ \ref{D(F)} 
show the dependence of the synchronization degree $D(F)$ for 
$\tau_a/T=0$, 1/3, and 2/3, where $T=2\pi/\Omega$ is the oscillation
period. Obviously, a significant information  loss occurs 
when $\tau_a$ becomes comparable to $T$, leading to a decrease of $D$. 
The curves $D(F)$ saturate at large $F$ allowing us to introduce 
the dependence $D_{max}(\tau )$. Calculations for the parameters of
Fig.\ \ref{D(F)} show pretty good synchronization, $D_{max} =0.993$, 
for $\tau_a=T/30$, while $D_{max} =0.98$, 0.92, and 0.57 
for $\tau_a =T/10$, $T/3$, and $2T/3$, respectively.

\begin{figure} [t] 
\centerline{
\epsfxsize=3.0in 
%\hspace{0.3cm}
\epsfbox{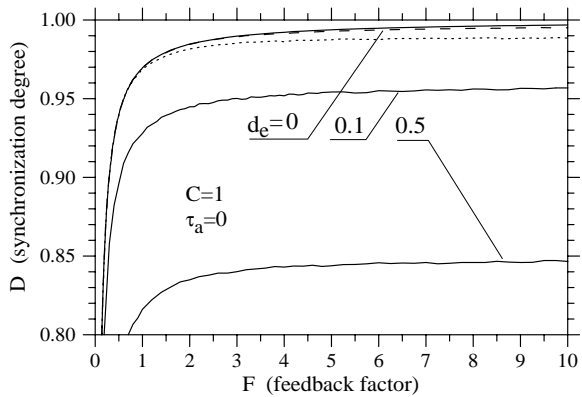}
%\epsfbox{qq.eps} 
} 
\vspace{0.4cm} 
\caption{Dependence $D(F)$ for ${\cal C}=1$, $\tau_a=0$, and several 
magnitudes of dephasing due to environment: $d_e=0$, 0.1, and 0.5. Dashed 
and dotted lines correspond to $d_e=0$ and limitation of $H_{fb}$ by 
0 and $H/2$, respectively. 
 }
\label{D(F)eta}\end{figure}

        The main potential practical importance of the quantum feedback 
is the ability to suppress the effect of the qubit dephasing caused by 
interaction with the environment (see Fig.\ \ref{schematic}). 
This can be used, for example, for qubit initialization in a solid-state
quantum computer. 
        Solid lines in Fig.\ \ref{D(F)eta} show the dependence 
$D(F)$ for several
magnitudes of the dephasing due to environment, $d_e=0$, 0.1, and 0.5, 
where $d_e\equiv \gamma_e /[(\Delta I)^2/4S_I]$ is the 
ratio between the qubit coupling to the environment and to the detector
(we still assume an ideal detector). First of all, we see that the feedback
still maintains the qubit phase synchronization for infinitely long time. 
However, for finite $d_e$ the degree of synchronization $D$ saturates 
at a level less than unity. We have studied numerically the 
dependence $D_{max}(d_e)$ for ${\cal C}=1/2$, 1, and 2 (while $\tau_a=0$ 
and $\eta =1$) and found a linear dependence at small
$d_e$: $D_{max}\simeq 1- 0.5\, d_e$. [A little better formula 
$D_{max}\simeq 1-0.5\, d_e/(1+d_e)$ works reasonably well up to 
$d_e \lesssim 1$.] 
  This means that the feedback loop can efficiently suppress the qubit 
dephasing due to the coupling to the environment if this coupling is much 
weaker than the qubit coupling to a nearly ideal detector.

        Notice that the solid lines shown in Figs.\ \ref{D(F)} and 
\ref{D(F)eta} are calculated assuming the feedback control of the
tunnel matrix element 
$H_{fb}=H[1-F\times \Delta \phi]$ even when $H_{fb}$ becomes negative
(this is also an assumption for the analytical results). To eliminate 
this unphysical assumption we have also performed numerical calculations 
with restrictions $H_{fb}>0$ and $H_{fb}>H/2$. 
This leads to rather minor modifications of the presented curves (dashed 
and dotted lines in Fig.\ \ref{D(F)eta} show the results for $d_e=0$ and 
$\tau_a=0$). However, important difference is that $D(F)$ goes down at
large $F$, so the optimum $D_{max}$ is achieved at some finite value of $F$.
More detailed study of this problem will be presented elsewhere. 

        Besides the discussed feedback based on $\Delta \phi$ calculation, 
we have also studied a ``direct'' feedback loop in which 
$H_{fb}(t)/H-1=F\{2[I_a(t)-I_0]/\Delta I -  \cos [\Omega (t-\tau_a/2)]\}
\, \sin [\Omega (t-\tau_a/2)]$
(we call it also a ``naive'' feedback because this control formula 
is easily designed 
from the naive assumption that the detector current directly follows the
evolution of $\rho_{11}$). 
Direct feedback is much simpler for experimental realization since
it does not require real-time solution of the Bayesian equations 
(direct feedback in quantum optics has been studied in Refs.\ 
\cite{Wiseman,Wiseman-94,Tombesi,Hofmann,Wang}). 
Quite surprisingly for us, the direct feedback can also provide 
a good phase 
synchronization of quantum oscillations if $F/{\cal C}$ is close to 1/4 
(see dotted line in Fig.\ \ref{D(F)}). 
 However, it requires more careful choice of $F$ and $\tau_a$ than for 
the Bayesian feedback, and also suffers more significantly from the 
restriction on $H_{fb}$ variation. 
The results in more detail will be discussed elsewhere.

Experimentally, besides the realization of quantum feedback control 
of a DQD continuously measured by a QPC, one can also think about the qubit
based on a single-Cooper-pair box measured by a single-electron transistor
(see discussion in \cite{Kor-meas2}). This realization can be preferable
because of a rapid progress of metallic single-electronics technology. 
However, the problems are high output impedance of the single-electron
transistor and its nonideality as a quantum detector. 
The third potential realization can be based on SQUIDs. For any realization
the major problem is the bandwidth: the feedback should be at least faster than
the qubit dephasing. Because of that, the quantum feedback of a solid-state 
qubit should probably be attempted only after the realization of recently 
proposed Bell-type two-detector correlation experiment,\cite{Kor-exp}
which would show the possibility of quantum monitoring, the first step to
quantum feedback control.

The authors thank A. J. Leggett and G. J. Milburn, H. M. Wiseman, and
A. C. Doherty for useful discussions. 
The work was supported by NSA and ARDA under ARO grant DAAD19-01-1-0491.


\begin{references}
\bibitem{decay} We neglect the possibility of a qubit decay, which 
	is important in optical setups but much less important in 
	solid-state realizations.
\bibitem{feedback-Doherty} The feedback considered in Ref.\ 
	\protect\cite{Doherty-3} is somewhat similar and in our notations
	would correspond to $H_{fb}=[1-F\times \mbox{sign}(\Delta \phi )] H$.
	Unfortunately, the dependence of the feedback loop operation on the 
	feedback strength was not studied in Ref.\ \protect\cite{Doherty-3}.
\end{references}
\end{document}